\documentclass{iau}
\usepackage{graphicx}

\newcommand{\ltsima}{$\; \buildrel < \over \sim \;$}
\newcommand{\simlt}{\lower.5ex\hbox{\ltsima}}
\newcommand{\gtsima}{$\; \buildrel > \over \sim \;$}
\newcommand{\simgt}{\lower.5ex\hbox{\gtsima}}

\newcommand{\cgs}{ ${\rm erg~cm}^{-2}~{\rm s}^{-1}$} 
\newcommand{\lum}{\rm erg~s$^{-1}$}

\def\lesssim{\mathrel{\hbox{\rlap{\hbox{\lower4pt\hbox{$\sim$}}}\hbox{$<$}}}}
\def\gtrsim{\mathrel{\hbox{\rlap{\hbox{\lower4pt\hbox{$\sim$}}}\hbox{$>$}}}}

\def\micron{\hbox{$\mu$m}}

\def\ab1450{$AB_{1450(1+z)}$}

\def\xray{\hbox{X-ray}}

\def\oiii{\hbox{[O\ {\sc iii}}]}

\def\nev{\hbox{[Ne\ {\sc v}}]}

\def\09104{IRAS~09104$+$4109}
\def\I09104{I09104}

\def\msun{M$_{\odot}$}
\def\edd_ratio{$\log\ L_{\rm bol}/L_{\rm Edd}$}

\def\l58{{$(\lambda L_{\lambda})_{\mbox{{\rm \scriptsize 5.8\micron}}}$}}
\def\lmir2{{$(\lambda L_{\lambda})_{\mbox{{\rm \scriptsize 12.3\micron}}}$}}
\def\s1{{S$_{\mbox{{\rm \scriptsize 3.6\micron}}}$}}
\def\irac2{{S$_{\mbox{{\rm \scriptsize 4.5\micron}}}$}}
\def\f3{{S$_{\mbox{{\rm \scriptsize 5.8\micron}}}$}}
\def\mic8{{S$_{\mbox{{\rm \scriptsize 8\micron}}}$}}
\def\f24{{F$_{\mbox{{\rm \scriptsize 24\micron}}}$}}

\def\alma{{\it ALMA\/}}
\def\iram{{\it IRAM\/}}
\def\herschel{{\it Herschel\/}}
\def\chandra{{\it Chandra\/}}

\def\spitzer{{\it Spitzer\/}}
\def\wise{{\it WISE\/}}

\def\nustar{{\it NuSTAR\/}}
\def\xmm{{XMM-{\it Newton\/}}}
\def\suzaku{{\it Suzaku\/}}

\def\swift{{\it Swift\/}}
\def\integral{{\it Integral\/}}
\def\nustar{{\it NuSTAR\/}}
\def\aj{\textit{AJ}}
\def\araa{\textit{ARA\&A}}
\def\apj{\textit{ApJ}}
\def\apjl{\textit{ApJ} (Letters)}
\def\apjs{\textit{ApJS} (Supplement)}
\def\aap{\textit{A\&A}}
\def\aapl{\textit{A\&A} (Letters)}
\def\mnras{\textit{MNRAS}}
\def\mnrasl{\textit{MNRAS} (Letters)}

\def\nature{\textit{Nature}}
\def\science{\textit{Science}}
\def\memsai{\textit{Mem.~Soc.~Astron.~Italiana}}

\title[Obscured accretion from AGN surveys]
{Obscured accretion from AGN surveys}

\author[Cristian Vignali]   
{Cristian Vignali$^{1,2}$}

\affiliation{$^1$Dipartimento di Fisica e Astronomia, Universit\`a di Bologna, 
Viale Berti Pichat 6/2, 40127 Bologna, Italy \\
$^2$INAF--Osservatorio Astronomico di Bologna, Via Ranzani 1, 40127 Bologna, 
Italy\\ email: {\tt cristian.vignali@unibo.it}} 

\pubyear{2014}
\volume{xxx}  
\pagerange{???--???}
\jname{IAU 304: Multiwavelegth AGN Surveys and Studies}
\editors{A. Mickaelian, F. Aharonian, D. Sanders Editor, eds.}
\begin{document}

\maketitle

\begin{abstract}
Recent models of super-massive black hole (SMBH) and host galaxy joint 
evolution predict the presence of a key phase where accretion, traced by 
obscured Active Galactic Nuclei (AGN) emission, is coupled with powerful 
star formation. Then feedback processes likely self-regulate the SMBH growth 
and quench the star-formation activity. 
AGN in this important evolutionary phase have been 
revealed in the last decade via surveys at different wavelengths. 
On the one hand, moderate-to-deep \xray\ surveys have allowed a systematic 
search for heavily obscured AGN, up to very high redshifts (z$\approx$5). 
On the other hand, infrared/optical surveys have been invaluable in offering 
complementary methods to select obscured AGN 
also in cases where the nuclear \xray\ emission below 10~keV is 
largely hidden to our view. 
In this review I will present my personal perspective of the field of obscured 
accretion from AGN surveys. 
\end{abstract}

\firstsection 
\section{Introduction}
One of the main science goals of modern observational cosmology is devoted to 
understand how galaxies and SMBHs at their centers grow together. 
Their close link leaves imprints in several relations observed in the local 
Universe between the mass of the black holes and the properties of the host 
galaxies (e.g., their velocity dispersion; 
\cite[Gebhardt et al. 2000]{gebhardt2000}; 
\cite[Ferrarese \& Merritt 2000]{ferrarese2000}). 
The emerging picture is that AGN are the key to understand 
the nature of such close connection, since the 
mass function of local SMBHs can be reasonably explained by the growth of seed 
black holes (whatever the origin of such seeds is) during AGN phases (e.g., 
\cite[Soltan 1982]{soltan1982}; \cite[Marconi et al. 2004]{marconi2004}). 

The entire picture, related to the so-called AGN-galaxy co-evolution scenario, 
has been presented in many works over the last decade, and has been perfectly 
synthesized in Fig.~1 of \cite{hopkins2008}, along the path traced by the 
original suggestion of \cite[Sanders et al. 1988]{sanders1988} (see also 
\cite[Sanders \& Mirabel 1996]{sanders1996}). 
Concisely, current quasar/host galaxy co-evolution models predict the 
existence of a dust-enshrouded phase associated with rapid SMBH growth and 
active star formation, largely triggered by multiple galaxy mergers and 
encounters (e.g., \cite[Silk \& Rees 1998]{silk1998}; 
\cite[Di Matteo et al. 2005]{dimatteo2005}; 
\cite[Menci et al. 2008]{menci2008}; \cite[Zubovas \& King 2012]
{zubovas2012}; \cite[Lamastra et al. 2013]{lamastra2013}). 
This phase is likely associated to obscured AGN growth in strongly 
star-forming (sub-millimeter) galaxies (e.g., 
\cite[Alexander et al. 2005]{alexander2005}). 
Finally, massive quasar-driven outflows blow away most of the cold gas 
reservoir, creating a population of ``red-and-dead'' gas-poor elliptical 
galaxies (e.g., \cite[Cattaneo et al. 2009]{cattaneo2009}). 

Support to this scenario comes from observations of wide-angle 
molecular outflows extending few kpc from the nucleus in some quasars hosted 
in ultra-luminous infrared galaxies; these systems, typically 
characterized by mass loss rates much larger than the ongoing star-formation 
rate (e.g., \cite[Feruglio et al. 2010]{feruglio2010}; 
\cite[Sturm et al. 2011]{sturm2011}; \cite[Rupke \& Veilleux 2013]{rupke2013}; 
\cite[Cicone et al. 2013]{cicone2013}), are observed up to very high redshifts 
(\cite[Maiolino et al. 2012]{maiolino2012}; 
\cite[Borguet et al. 2013]{borguet2013}). 
Similarly, observations of powerful outflows in neutral and ionized gas have 
also been collected over the past few years 
(e.g., \cite[Nesvadba et al. 2008]{nesvadba2008}; 
\cite[Alexander et al. 2010]{alexander10}; 
\cite[Harrison et al. 2012]{harrison2012}). 
This feedback process ascribed to quasars is most certainly related to 
radiation-driven winds and is often invoked to explain why SMBHs and galaxies 
stop growing at a certain phase of their life; for a more comprehensive 
discussion on this issue, see the review by C.M. Harrison (this Volume). 
Evidences for ultra-fast outflows (with velocities typically up to 
0.1--0.4c) have been recently observed in X-rays in a sizable sample of AGN, 
both in the local Universe (e.g., 
\cite[Tombesi et al. 2010, 2011, 2012]{tombesi2010,tombesi2011,tombesi2012}; 
\cite[Gofford et al. 2013]{gofford2013}; 
\cite[Reeves et al. 2003]{reeves2003}) and at high redshift (e.g., 
\cite[Chartas et al. 2002, 2007]{chartas2002,chartas2007}; 
\cite[Saez et al. 2009]{saez2009}). 
The connection between molecular and highly ionized gas is, however, from from 
being assessed, and will constitute undoubtedly one of the prime science goals 
of the coming years using \alma\ and \iram\ facilities at long wavelengths 
and \chandra\ and \xmm\ in the \xray\ domain.

According to the scenario described above, the main trigger mechanism 
of BH accretion and growth is ascribed to galaxy mergers and interactions, 
at least in the most luminous and massive systems. 
Most of their mass is assembled in short periods ($\approx$10--100~Myr) of 
``bursting'' nuclear and star-forming activity, while the bulk of galaxies 
and SMBHs grow their mass in a secular (i.e., ``smooth'') mode over timescales 
of Gyrs (e.g., \cite[Daddi et al. 2007a; Hickox et al. 2009]
{daddi2007a,hickox2009}). 
This picture has recently been confirmed by \herschel\ surveys, 
showing a distinction between the bulk of galaxies growing 
quietly (in the so-called ``main sequence'') and the minority of the galaxy 
population whose growth happens mostly during events of mergers of gas-rich 
galaxies in the so-called ``starburst mode'' (e.g., 
\cite[Elbaz et al. 2011]{elbaz2011}; 
\cite[Rodighiero et al. 2011]{rodighiero2011}; see also \cite[Rosario et al. 
2013]{rosario2013}). 

As a natural consequence of the merger scenario, a key phase 
in the AGN and galaxy life is when large amounts of gas are funneled to 
the center, thus inducing both obscured accretion and star formation (e.g., 
\cite[Treister et al. 2010]{treister2010}). 
Significant efforts have been made recently to search for and characterize, 
as much as possible, the most heavily obscured AGN and 
quasars, dubbed Compton thick, characterized by column densities above 
$1.5\times10^{24}$~cm$^{-2}$ (see \cite[Comastri 2004]{comastri2004} for 
a review); such absorbers strongly limit the possibility for these sources 
of being detected at energies below 10~keV (where sensitive \xray\ imaging 
instruments are currently operative). Therefore, in order to provide a census 
as complete as possible of this source population, a multi-wavelength 
synergistic approach is needed. 

In this review I will focus on some aspects and methods of investigation 
that I think are important in the quest for heavily obscured AGN. As such, 
this proceeding is not meant to provide an exhaustive view of this topic. 
Further and, possibly, alternative approaches in this research field 
and consequences for AGN synthesis models for the \xray\ background (XRB) are 
addressed by other authors in this Volume (e.g., A. Barger, A. Del Moro, 
S. Juneau, A. Levenson, S. Mateos, L. Spinoglio, D. Stern, E. Treister, 
Y. Ueda).

\section{Searching for heavily obscured AGN}
The problem of finding heavily obscured AGN and quasars can be tackled 
following various prescriptions and adopting different approaches. 
The bad news is that there is no way to obtain a complete census of this 
AGN population either using single-band observations or a unique selection 
method/criterion. The good news is that the multi-wavelength observing 
campaigns which characterize most of the current surveys offer a unique 
possibility to detect the most obscured AGN, up to very high redshifts. 
Adopting several selection criteria and keeping in mind the observational 
biases intrinsic to each detection band are what we need in the future to 
infer the demographics of these elusive AGN and use them to provide 
``boundary'' conditions and useful constraints 
to AGN/galaxy co-evolution models. 

\begin{figure}[t]
\begin{center}
\includegraphics[width=0.65\textwidth,angle=90]{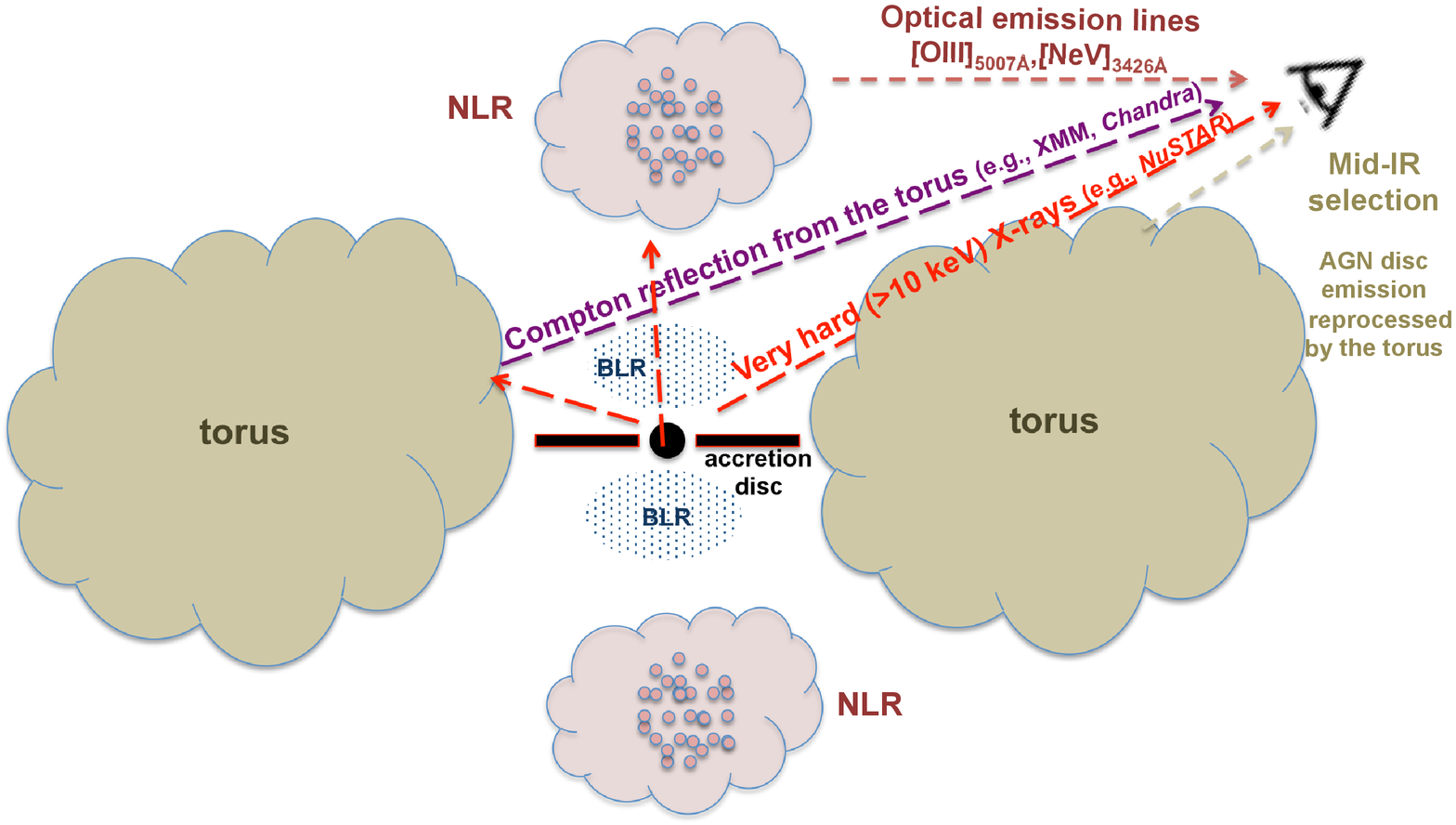}
\caption{Schematic view of AGN (not in scale). Emphasis is given to the 
emission components which, at different wavelengths, allow for the detection of 
obscured AGN. BLR and NLR stand for broad-line region and narrow-line region, 
respectively.}
\label{agn_schematic_view}
\end{center}
\end{figure}

In the following, I will try to elucidate some detection techniques adopted 
to find obscured AGN, which are schematized in Fig.~\ref{agn_schematic_view}. 
In particular, I am referring to methods related to \xray\ 
($\S$\ref{xray_surveys}), mid-infrared (mid-IR; $\S$\ref{mid_ir}) and optical 
selection ($\S$\ref{optical}). 

\subsection{Hard X-ray surveys}
\label{xray_surveys}
According to the unified model for AGN (\cite[Antonucci 1993]{antonucci1993}), 
the \xray\ emission, once it intercepts the obscuring material 
(i.e., the torus; see Fig.~\ref{agn_schematic_view}), can 
be profoundly depressed in the \xray\ band. In particular, 
if the optical depth for Compton scattering ($\tau=N_H\times\sigma_T$) does 
not exceed values of the order of ``{\it a few}", \xray\ photons with energies 
higher than 10--15~keV are able to penetrate the obscuring material and reach 
the observer. For higher values of $\tau$, the entire \xray\ spectrum is 
depressed by Compton down-scattering and the \xray\ photons are effectively 
trapped by the obscuring material irrespective of their energy. 
The former class of sources (mildly Compton thick) can be efficiently 
detected by \xray\ instruments above 10~keV, while for the 
latter (heavily Compton thick) their nature may be inferred through indirect 
arguments, such as the presence of a strong iron K$\alpha$ emission line over a 
flat reflected continuum. 
Mildly Compton-thick AGN are the most promising candidates to explain 
the residual (i.e., not resolved yet) spectrum of the cosmic XRB at 
its 30~keV peak (e.g., \cite[Worsley et al. 2005; Gilli et al. 2007; 
Ballantyne 2009; Treister et al. 2009; Moretti et al. 2012; Shi et al. 2013]
{worsley2005; gilli2007; ballantyne2009; treister2009; moretti2012; shi2013}) 
but only a handful of them are known 
(i.e., have been classified as such beyond any reasonable doubt) 
outside the local Universe (e.g., \cite[Iwasawa et al. 2005]{iwasawa2005}). 
An unbiased census of extremely obscured AGN would require to survey 
the hard \xray\ above 10~keV with a fairly good sensitivity. A step 
forward in this direction is being provided by the \swift/BAT and 
\integral/IBIS surveys (e.g., \cite[Tueller et al. 2008; 
Beckmann et al. 2009; Vasudevan et al. 2013]{tueller2008; beckmann2009; 
vasudevan2013}), which have covered a large portion of the sky though 
limited to relatively bright \xray\ fluxes ($\approx10^{-11}$~\cgs), hence 
to low redshifts, and have resolved less than 10\% of the XRB. 
The spectral characterization of the heavily obscured AGN discovered 
in these shallow hard \xray\ surveys often required follow-up observations 
with the more sensitive instruments onboard \chandra, \xmm\ and \suzaku\ 
(e.g., \cite[Eguchi et al. 2009; Comastri et al. 2010; Winter et al. 2010; 
Severgnini et al. 2011; Burlon et al. 2011]{eguchi2009; comastri2010;
severgnini2011; winter2010; burlon2011}); this approach led to an estimate 
of a fraction of $\approx$~10--20\% of Compton-thick AGN among hard \xray\ 
selected samples (e.g., \cite[Malizia et al. 2009; Burlon et al. 2011; 
Vasudevan et al. 2013]{malizia2009; burlon2011; vasudevan2013}). 
Data from the \nustar\ satellite, having imaging capabilities 
up to $\approx$~80~keV, can shed new light on this topic at sensitivities 
more than a factor 100 better than those achieved by \integral\ and \swift\ 
(\cite[Alexander et al. 2013]{alexander2013}). 

Deep \xray\ surveys with sensitive imaging instruments (\chandra\ and \xmm) 
can push the detection of Compton-thick AGN at considerably higher 
redshifts (e.g., $z=4.75$, \cite[Gilli et al. 2011]{gilli2011}). 
Indications of Compton-thick material in AGN and quasars have been found 
by many authors, often coupled to powerful star formation 
(from few hundred to $\approx$~1000 \msun/yr), mostly using the deep exposures 
in the \chandra\ Deep Field-South (CDF-S) provided by both \chandra\ (currently 
4~Ms -- \cite[Xue et al. 2011]{xue2011} -- close to be extended to 7~Ms) 
and \xmm\ ($\approx$~3~Ms; \cite[Ranalli et al. 2013]{ranalli2013}); see, e.g., 
\cite[Tozzi et al. (2006); Georgantopoulos et al. (2009, 2013); 
Comastri et al. (2011); Feruglio et al. (2011); Brightman \& Ueda (2012); 
Vito et al. (2013)]
{tozzi2006; georgantopoulos2009; georgantopoulos2013; comastri2011; 
feruglio2011; brightman2012; vito2013}. 
For a significant fraction of \xray\ sources found in deep fields, the 
signal-to-noise ratio of the spectra is limited and does not allow for 
a proper characterization of the source spectral complexities. 
Further constraints on the obscured AGN population may be derived 
using \xray\ stacking techniques which take benefit of the good spatial 
resolution (primarily offered by \chandra) and allow exploration of 
considerably deeper \xray\ fluxes (e.g., \cite[Xue et al. 2012]{xue2012}). 
However, even the deepest \xray\ exposures currently available miss a 
significant number of very obscured AGN, hence a not negligible fraction 
of the accretion power in the Universe. 

Another interesting result which is emerging from deep \xray\ surveys is 
related to the increasing fraction of heavily obscured quasars from z=0 
to z$\approx$3--4; a similar trend is apparently not observed in lower 
luminosity AGN (\cite[Iwasawa et al. 2012; Vito et al. 2013]
{iwasawa2012; vito2013}). 
Since the fraction of AGN in mergers seems to increase with the bolometric 
luminosity (\cite[Treister et al. 2012]{treister2012}), we may expect 
that at high redshift, when the merger rate was higher, a larger gas fraction 
(producing obscuration) was available in galaxies. The planned 
extension of \chandra\ observations in the CDF-S, coupled to very deep infrared 
data (e.g., CANDELS), will hopefully allow us to explore this hypothesis 
at very high redshifts in a couple of years.

\subsection{Mid-infrared selection}
\label{mid_ir}
The mid-IR regime offers much potential for discovery of heavily obscured 
AGN, since any primary AGN continuum (i.e., disc emission) 
that is absorbed must ultimately come out 
at these wavelengths after being thermally reprocessed by the torus 
(see Fig.~\ref{agn_schematic_view}). 
Thus, sources with weak emission in the optical band (because of extinction) 
and relatively bright mid-IR emission can be counted as heavily obscured 
AGN candidates, unless a significant contribution in the mid-IR comes 
from star-formation processes (PAH features and continuum emission). 
This probably ``basic'' high mid-IR/optical flux-ratio selection method
found support in many works in the era of the \spitzer\ observatory 
(e.g., \cite[Mart{\'{\i}}nez-Sansigre et al. 2005; Houck et al. 2005; 
Weedman et al. 2006]{sansigre2005; houck2005; weedman2006}), 
and allowed \cite{dey2008} 
to define a new class of sources at $z\approx$~2, the 
dusty obscured galaxies (DOGs), having $F_{24\micron}/F_{R}>1000$. 
Among these, we may expect some of the most obscured AGN, especially if 
a selection at $F_{24\micron}>1$~mJy is adopted to limit the contamination 
from star-forming galaxies (e.g., \cite[Sacchi et al. 2009]
{sacchi2009}). 
This selection is different from those allowed by the widely adopted 
mid-IR color-color diagrams (e.g., \cite[Lacy et al. 2004; 
Stern et al. 2005; see also Donley et al. 2012]
{lacy2004; stern2005; donley2012}), 
where separating the most heavily obscured AGN from the remaining source 
populations is not a trivial job (e.g., \cite[Castell{\'o}-Mor et al. 2013]
{castello_mor2013}). 
However, only \xray\ data have been able to provide the smoking gun of the 
truly Compton-thick nature for a fraction of the high mid-IR/optical flux-ratio 
sources (e.g., \cite[Polletta et al. 2006; Lanzuisi et al. 2009; 
Georgantopoulos et al. 2011; see also Severgnini et al. 2012]
{polletta2006; lanzuisi2009; georgantopoulos2011; severgnini2012}). 
Furthermore, \xray\ stacking analyses have allowed to place observational 
constraints, for the first time, to the space density of Compton-thick AGN at 
high redshifts ($z\approx2-3$; \cite[Daddi et al. 2007b; Fiore et al. 2008, 
2009; Bauer et al. 2010; Alexander et al. 2011; but see also Georgakakis 
et al. 2010]{daddi2007b; fiore2008; fiore2009; bauer2010; 
alexander2011; georgakakis2010}). 

Extension of the mid-IR search for heavily obscured AGN is within the 
capabilities offered by \wise, as shown by \cite{mateos2013} and D. Stern 
(this Volume).

\subsection{Optical selection}
\label{optical}
The selection of obscured AGN using optical spectroscopy proceeds primarily 
through the detection of high-ionization emission lines, e.g., 
\oiii5007\AA\ and \nev3426\AA. 
These lines, being produced in the narrow-line region (NLR), do not 
suffer from extinction from the torus and are considered good proxies of the 
nuclear intrinsic power (see Fig.~\ref{agn_schematic_view}). 
Applying the relation between \oiii\ and 2--10~keV emission 
(e.g., \cite[Mulchaey et al. 1994; Heckman et al. 2005; 
Panessa et al. 2006]{mulchaey1994; heckman2005; panessa2006}) to the sample 
of narrow-line AGN from the Sloan Digital Sky Survey of 
\cite{zakamska2003} led some authors (e.g., \cite[Vignali et al. 2006, 
2010; Ptak et al. 2006]{vignali2006; vignali2010; ptak2006}) to the discovery 
of about a dozen of Compton-thick AGN candidates. 
These studies allowed a first estimate of the space density of this 
obscured AGN population at $z\approx0.3-0.8$. According to 
\cite[Gilli et al. (2007, 2013)]{gilli2007; gilli2013} XRB models, 
the fraction of XRB emission at 20~keV produced by Compton-thick AGN 
and still ``missing'' has a peak at $z\approx0.7$ 
and is mostly due to Seyfert-like objects, with intrinsic 2--10~keV 
luminosity below $10^{44}$~\lum. Moving these investigations to slightly 
higher redshifts requires the use of the \nev\ emission line, which has the 
advantage of being an unambiguous marker of AGN (with a ionization potential 
of 97~eV vs. 54~eV of \oiii) but is $\approx$~9 times weaker than \oiii\ 
and suffers from stronger extinction. 
Calibrating the \xray-to-\nev\ luminosity ratio on a sample of local AGN, 
\cite{gilli2010} show that values $<15$ are highly indicative of Compton-thick 
obscuration. How effective this line is in finding Compton-thick AGN has been 
recently confirmed by \cite{mignoli2013}, where narrow-line AGN were selected 
from the zCOSMOS survey and \xray\ coverage was provided by \chandra\ 
(\cite[Vignali et al., in preparation]{vignali2014}). About 40\% of 
the original $\approx70$ candidates are consistent with being Compton thick 
(in line with \cite[Gilli et al. 2007]{gilli2007} model). 
We note, however, that optical spectroscopy, because of extinction within the 
NLR, is far from offering a complete census of obscured 
AGN (see $\S$3.3 of \cite[Mignoli et al. 2013]{mignoli2013}). 
Further insights into the properties of these \nev-selected Compton-thick 
AGN will come out by using their mid-IR emission as another proxy of the 
nuclear emission (e.g., \cite[Gandhi et al. 2009]{gandhi2009}) to be compared 
to the observed \xray\ luminosity.

\section{Conclusions}
Obscured AGN growth is a key phase in SMBH/galaxy co-evolution models. 
As the census of such objects is difficult, especially at high redshifts, 
a multi-wavelength synergistic approach is needed, requiring deep \xray\ 
exposure, mid-IR data and, possibly, optical/near-IR spectroscopy. 
Whatever the adopted selection method is, X-rays represent a powerful 
and fundamental probe through direct \xray\ spectroscopy and stacking analysis. 

\acknowledgments
I would like to thank the organizers of this Symposium and all of my close 
collaborators for the many enlightening discussions.


\begin{thebibliography}{}

\bibitem[Alexander et al. (2005)]{alexander2005} 
{Alexander, D.~M., Smail, I., Bauer, F.~E., et al.} 2005, 
\nature, 434, 738 

\bibitem[Alexander et al. (2010)]{alexander2010} 
{Alexander, D.~M., Swinbank, A.~M., Smail, I., et al.} 2010, 
\mnras, 402, 2211 

\bibitem[Alexander et al. (2011)]{alexander2011} 
{Alexander, D.~M., Bauer, F.~E., Brandt, W.~N., et al.} 2011, 
\apj, 738, 44 

\bibitem[Alexander et al. (2013)]{alexander2013} 
{Alexander, D.~M., Stern, D., Del Moro, A., et al.} 2013, 
\apj, 773, 125 

\bibitem[Antonucci (1993)]{antonucci1993} 
{Antonucci, R.} 1993, 
\araa, 31, 473 

\bibitem[Ballantyne (2009)]{ballantyne2009} 
{Ballantyne, D.~R.} 2009, 
\apj, 698, 1033 

\bibitem[Bauer et al. (2010)]{bauer2010} 
{Bauer, F.~E., Yan, L., Sajina, A., \& Alexander, D.~M.} 2010, 
\apj, 710, 212 

\bibitem[Beckmann et al. (2009)]{beckmann2009} 
{Beckmann, V., Soldi, S., Ricci, C., et al.} 2009, 
\aap, 505, 417 

\bibitem[Borguet et al. (2013)]{borguet2013}
{Borguet, B.~C.~J., Arav, N., Edmonds, D., Chamberlain, C., \& Benn, C.} 2013, 
\apj, 762, 49 


\bibitem[Brightman \& Ueda (2012)]{brightman2012} 
{Brightman, M., \& Ueda, Y.} 2012, 
\mnras, 423, 702 

\bibitem[Burlon et al. (2011)]{burlon2011} 
{Burlon, D., Ajello, M., Greiner, J., et al.} 2011, 
\apj, 728, 58 

\bibitem[Castell{\'o}-Mor et al. (2013)]{castello_mor2013} 
{Castell{\'o}-Mor, N., Carrera, F.~J., Alonso-Herrero, A., et al.} 2013, 
\aap, 556, A114 

\bibitem[Cattaneo et al. (2009)]{cattaneo2009} 
{Cattaneo, A., Faber, S.~M., Binney, J., et al.} 2009, 
\nature, 460, 213 

\bibitem[Chartas et al. (2002)]{chartas2002} 
{Chartas, G., Brandt, W.~N., Gallagher, S.~C., \& Garmire, G.~P.} 2002, 
\apj, 579, 169 

\bibitem[Chartas et al. (2007)]{chartas2007} 
{Chartas, G., Eracleous, M., Dai, X., Agol, E., \& Gallagher, S.} 2007, 
\apj, 661, 678 

\bibitem[Cicone et al. (2013)]{cicone2013} 
{Cicone, C., Maiolino, R., Sturm, E., et al.} 2013, 
\aap, in press (arXiv:1311.2595)

\bibitem[Comastri (2004)]{comastri2004} 
{Comastri, A.} 2004, in: A.J. Barger (eds.), 
\textit{Supermassive Black Holes in the Distant Universe}, 
(Astrophysics and Space Science Library: Kluwer), 308, 245 

\bibitem[Comastri et al. (2010)]{comastri2010} 
{Comastri, A., Iwasawa, K., Gilli, R., et al.} 2010, 
\apj, 717, 787 

\bibitem[Comastri et al. (2011)]{comastri2011} 
{Comastri, A., Ranalli, P., Iwasawa, K., et al.} 2011, 
\aap, 526, L9 

\bibitem[Daddi et al. (2007a)]{daddi2007a} 
{Daddi, E., et al.} 2007a, 
\apj, 670, 156 

\bibitem[Daddi et al. (2007b)]{daddi2007b} 
{Daddi, E., et al.} 2007b, 
\apj, 670, 173 


\bibitem[Dey et al. (2008)]{dey2008} 
{Dey, A., Soifer, B.~T., Desai, V., et al.} 2008, 
\apj, 677, 943 

\bibitem[Di Matteo et al. (2005)]{dimatteo2005} 
{Di Matteo, T., Springel, V., \& Hernquist, L.} 2005, 
\nature, 433, 604 

\bibitem[Donley et al. (2012)]{donley2012} 
{Donley, J.~L., Koekemoer, A.~M., Brusa, M., et al.} 2012, 
\apj, 748, 142 

\bibitem[Eguchi et al. (2009)]{eguchi2009} 
{Eguchi, S., Ueda, Y., Terashima, Y., Mushotzky, R., \& Tueller, J.} 2009, 
\apj, 696, 1657
 
\bibitem[Elbaz et al. (2011)]{elbaz2011} 
{Elbaz, D., Dickinson, M., Hwang, H.~S., et al.} 2011, 
\aap, 533, A119 

\bibitem[Ferrarese \& Merritt (2000)]{ferrarese2000} 
{Ferrarese, L., \& Merritt, D.} 2000, 
\apjl, 539, L9 

\bibitem[Feruglio et al. (2010)]{feruglio2010} 
{Feruglio, C., Maiolino, R., Piconcelli, E., et al.} 2010, 
\aapl, 518, L155 

\bibitem[Feruglio et al. (2011)]{feruglio2011} 
{Feruglio, C., Daddi, E., Fiore, F., et al.} 2011, 
\apjl, 729, L4 

\bibitem[Fiore et al. (2008)]{fiore2008} 
{Fiore, F., Grazian, A., Santini, P., et al.} 2008, 
\apj, 672, 94 

\bibitem[Fiore et al. (2009)]{fiore2009} 
{Fiore, F., Puccetti, S., Brusa, M., et al.} 2009, 
\apj, 693, 447 

\bibitem[Gandhi et al. (2009)]{gandhi2009} 
{Gandhi, P., Horst, H., Smette, A., et al.} 2009, 
\aap, 502, 457 

\bibitem[Gebhardt et al. (2000)]{gebhardt2000} 
{Gebhardt, K., Bender, R., Bower, G., et al.} 2000, 
\apjl, 539, L13 

\bibitem[Georgakakis et al. (2010)]{georgakakis2010} 
{Georgakakis, A., Rowan-Robinson, M., Nandra, K., et al.} 2010, 
\mnras, 406, 420 

\bibitem[Georgantopoulos et al. (2009)]{georgantopoulos2009} 
{Georgantopoulos, I., Akylas, A., Georgakakis, A., \& Rowan-Robinson, M.} 2009, 
\aap, 507, 747 

\bibitem[Georgantopoulos et al. (2011)]{georgantopoulos2011} 
{Georgantopoulos, I., Rovilos, E., Akylas, A., et al.} 2011, 
\aap, 534, A23 

\bibitem[Georgantopoulos et al. (2013)]{georgantopoulos2013} 
{Georgantopoulos, I., Comastri, A., Vignali, C., et al.} 2013, 
\aap, 555, A43 

\bibitem[Gilli et al. (2007)]{gilli2007} 
{Gilli, R., Comastri, A., \& Hasinger, G.} 2007, 
\aap, 463, 79 

\bibitem[Gilli et al. (2010)]{gilli2010} 
{Gilli, R., Vignali, C., Mignoli, M., et al.} 2010, 
\aap, 519, A92 

\bibitem[Gilli et al. (2011)]{gilli2011} 
{Gilli, R., Su, J., Norman, C., et al.} 2011, 
\apjl, 730, L28 

\bibitem[Gilli(2013)]{gilli2013} 
{Gilli, R.} 2013, 
\memsai, 84, 647 

\bibitem[Gofford et al. (2013)]{gofford2013} 
{Gofford, J., Reeves, J.~N., Tombesi, F., et al.} 2013, 
\mnras, 430, 60 

\bibitem[Harrison et al. (2012)]{harrison2012} 
{Harrison, C.~M., Alexander, D.~M., Swinbank, A.~M., et al.} 2012, 
\mnras, 426, 1073 

\bibitem[Heckman et al. (2005)]{heckman2005} 
{Heckman, T.~M., Ptak, A., Hornschemeier, A., \& Kauffmann, G.} 2005, 
\apj, 634, 161 

\bibitem[Hickox et al. (2009)]{hickox2009} 
{Hickox, R.~C., Jones, C., Forman, W.~R., et al.} 2009, 
\apj, 696, 891 

\bibitem[Hopkins et al. (2008)]{hopkins2008} 
{Hopkins, P.~F., Hernquist, L., Cox, T.~J., \& Kere{\v s}, D.} 2008, 
\apjs, 175, 356 

\bibitem[Houck et al. (2005)]{houck2005} 
{Houck, J.~R., Soifer, B.~T., Weedman, D., et al.} 2005, 
\apjl, 622, L105 

\bibitem[Iwasawa et al. (2005)]{iwasawa2005} 
{Iwasawa, K., Crawford, C.~S., Fabian, A.~C., \& Wilman, R.~J.} 2005, 
\mnras, 362, L20 

\bibitem[Iwasawa et al. (2012)]{iwasawa2012} 
{Iwasawa, K., Gilli, R., Vignali, C., et al.} 2012, 
\aap, 546, A84 

\bibitem[Lacy et al. (2004)]{lacy2004} 
{Lacy, M., Storrie-Lombardi, L.~J., Sajina, A., et al.} 2004, 
\apjs, 154, 166 


\bibitem[Lamastra et al. (2013)]{lamastra2013} 
{Lamastra, A., Menci, N., Fiore, F., et al.} 2013, 
\aap, 559, A56 

\bibitem[Lanzuisi et al. (2009)]{lanzuisi2009} 
{Lanzuisi, G., Piconcelli, E., Fiore, F., et al.} 2009, 
\aap, 498, 67 


\bibitem[Maiolino et al. (2012)]{maiolino2012} 
{Maiolino, R., Gallerani, S., Neri, R., et al.} 2012, 
\mnrasl, 425, L66 

\bibitem[Malizia et al. (2009)]{malizia2009} 
{Malizia, A., Stephen, J.~B., Bassani, L., et al.} 2009, 
\mnras, 399, 944 

\bibitem[Marconi et al. (2004)]{marconi2004} 
{Marconi, A., Risaliti, G., Gilli, R., et al.} 2004, 
\mnras, 351, 169 

\bibitem[Mart{\'{\i}}nez-Sansigre et al. (2005)]{sansigre2005} 
{Mart{\'{\i}}nez-Sansigre, A., Rawlings, S., Lacy, M., et al.} 2005, 
\nature, 436, 666 

\bibitem[Mateos et al. (2013)]{mateos2013} 
{Mateos, S., Alonso-Herrero, A., Carrera, F.~J., et al.} 2013, 
\mnras, 434, 941 

\bibitem[Menci et al. (2008)]{menci2008} 
{Menci, N., Fiore, F., Puccetti, S., \& Cavaliere, A.} 2008, 
\apj, 686, 219 

\bibitem[Mignoli et al. (2013)]{mignoli2013} 
{Mignoli, M., Vignali, C., Gilli, R., et al.} 2013, 
\aap, 556, A29 

\bibitem[Moretti et al. (2012)]{moretti2012} 
{Moretti, A., Vattakunnel, S., Tozzi, P., et al.} 2012, 
\aap, 548, A87 

\bibitem[Mulchaey et al. (1994)]{muclahey1994} 
{Mulchaey, J.~S., Koratkar, A., Ward, M.~J., et al.} 1994, 
\apj, 436, 586 

\bibitem[Nesvadba et al. (2008)]{nesvadba2008} 
{Nesvadba, N.~P.~H., Lehnert, M.~D., De Breuck, C., et al.} 2008, 
\aap, 491, 407 

\bibitem[Panessa et al. (2006)]{panessa2006} 
{Panessa, F., Bassani, L., Cappi, M., et al.} 2006, 
\aap, 455, 173 

\bibitem[Polletta et al. (2006)]{polletta2006} 
{Polletta, M.~d.~C., Wilkes, B.~J., Siana, B., et al.} 2006, 
\apj, 642, 673 

\bibitem[Ptak et al. (2006)]{ptak2006} 
{Ptak, A., Zakamska, N.~L., Strauss, M.~A., et al.} 2006, 
\apj, 637, 147 

\bibitem[Ranalli et al. (2013)]{ranalli2013} 
{Ranalli, P., Comastri, A., Vignali, C., et al.} 2013, 
\aap, 555, A42 

\bibitem[Reeves et al. (2003)]{reeves2003} 
{Reeves, J.~N., O'Brien, P.~T., \& Ward, M.~J.} 2003, 
\apjl, 593, L65 

\bibitem[Rodighiero et al. (2011)]{rodighiero2011} 
{Rodighiero, G., Daddi, E., Baronchelli, I., et al.} 2011, 
\apjl, 739, L40 

\bibitem[Rosario et al. (2013)]{rosario2013} 
{Rosario, D.~J., Trakhtenbrot, B., Lutz, D., et al.} 2013, 
\aap, 560, A72 

\bibitem[Rupke \& Veilleux (2013)]{rupke2013} 
{Rupke, D.~S.~N., \& Veilleux, S.} 2013, 
\apjl, 775, L15 

\bibitem[Sacchi et al. (2009)]{sacchi2009} 
{Sacchi, N., La Franca, F., Feruglio, C., et al.} 2009, 
\apj, 703, 1778 

\bibitem[Saez et al. (2009)]{saez2009} 
{Saez, C., Chartas, G., \& Brandt, W.~N.} 2009, 
\apj, 697, 194 

\bibitem[Sanders et al. (1988)]{sanders1988} 
{Sanders, D.~B., Soifer, B.~T., Elias, J.~H., et al.} 1988, 
\apj, 325, 74 

\bibitem[Sanders \& Mirabel (1996)]{sanders1996} 
{Sanders, D.~B., \& Mirabel, I.~F.} 1996, 
\araa, 34, 749 

\bibitem[Severgnini et al. (2011)]{severgnini2011} 
{Severgnini, P., Caccianiga, A., Della Ceca, R., et al.} 2011, 
\aap, 525, A38 

\bibitem[Severgnini et al. (2012)]{severgnini2012} 
{Severgnini, P., Caccianiga, A., \& Della Ceca, R.} 2012, 
\aap, 542, A46 

\bibitem[Shi et al. (2013)]{shi2013} 
{Shi, Y., Helou, G., \& Armus, L.} 2013, 
\apj, 777, 6 

\bibitem[Silk \& Rees (1998)]{silk1998} 
{Silk, J., \& Rees, M.~J.} 1998, 
\aapl, 331, L1 

\bibitem[Soltan (1982)]{soltan1982} 
{Soltan, A.} 1982, 
\mnras, 200, 115 

\bibitem[Stern et al. (2005)]{stern2005} 
{Stern, D., Eisenhardt, P., Gorjian, V., et al.} 2005, 
\apj, 631, 163 

\bibitem[Sturm et al. (2011)]{sturm2011} 
{Sturm, E., Gonz{\'a}lez-Alfonso, E., Veilleux, S., et al.} 2011, 
\apjl, 733, L16 

\bibitem[Tombesi et al. (2010)]{tombesi2010} 
{Tombesi, F., Cappi, M., Reeves, J.~N., et al.} 2010, 
\aap, 521, A57 

\bibitem[Tombesi et al. (2011)]{tombesi2011} 
{Tombesi, F., Cappi, M., Reeves, J.~N., et al.} 2011, 
\apj, 742, 44 

\bibitem[Tombesi et al. (2012)]{tombesi2012} 
{Tombesi, F., Cappi, M., Reeves, J.~N., \& Braito, V.} 2012, 
\mnrasl, 422, L1 

\bibitem[Tozzi et al. (2006)]{tozzi2006} 
{Tozzi, P., Gilli, R., Mainieri, V., et al.} 2006, 
\aap, 451, 457 

\bibitem[Treister et al. (2009)]{treister2009} 
{Treister, E., Urry, C.~M., \& Virani, S.} 2009, 
\apj, 696, 110 

\bibitem[Treister et al. (2010)]{treister2010} 
{Treister, E., Natarajan, P., Sanders, D.~B., et al.} 2010, 
\science, 328, 600 

\bibitem[Treister et al. (2012)]{treister2012} 
{Treister, E., Schawinski, K., Urry, C.~M., \& Simmons, B.~D.} 2012, 
\apjl, 758, L39 

\bibitem[Tueller et al. (2008)]{tueller2008} 
{Tueller, J., Mushotzky, R.~F., Barthelmy, S., et al.} 2008, 
\apj, 681, 113 

\bibitem[Vasudevan et al. (2013)]{vasudevan2013} 
{Vasudevan, R.~V., Brandt, W.~N., Mushotzky, R.~F., et al.} 2013, 
\apj, 763, 111 

\bibitem[Vignali et al. (2006)]{vignali2006} 
{Vignali, C., Alexander, D.~M., \& Comastri, A.} 2006, 
\mnras, 373, 321 

\bibitem[Vignali et al. (2010)]{vignali2010} 
{Vignali, C., Alexander, D.~M., Gilli, R., \& Pozzi, F.} 2010, 
\mnras, 404, 48 

\bibitem[Vignali et al. (in prep.)]{vignali2014} 
{Vignali, C., Mignoli, M., Gilli, R., et al.}, 
\aap, in preparation

\bibitem[Vito et al. (2013)]{vito2013} 
{Vito, F., Vignali, C., Gilli, R., et al.} 2013, 
\mnras, 428, 354 

\bibitem[Weedman et al. (2006)]{weedman2006} 
{Weedman, D.~W., Soifer, B.~T., Hao, L., et al.} 2006, 
\apj, 651, 101 

\bibitem[Winter et al. (2010)]{winter2010} 
{Winter, L.~M., Lewis, K.~T., Koss, M., et al.} 2010, 
\apj, 710, 503 

\bibitem[Worsley et al. (2005)]{worsley2005} 
{Worsley, M.~A., Fabian, A.~C., Bauer, F.~E., et al.} 2005, 
\mnras, 357, 1281 

\bibitem[Xue et al. (2011)]{xue2011} 
{Xue, Y.~Q., Luo, B., Brandt, W.~N., et al.} 2011, 
\apjs, 195, 10 

\bibitem[Xue et al. (2012)]{xue2012} 
{Xue, Y.~Q., Wang, S.~X., Brandt, W.~N., et al.} 2012, 
\apj, 758, 129 

\bibitem[Zakamska et al. (2003)]{zakamska2003} 
{Zakamska, N.~L., Strauss, M.~A., Krolik, J.~H., et al.} 2003, 
\aj, 126, 2125 

\bibitem[Zubovas \& King (2012)]{zubovas2012} 
{Zubovas, K., \& King, A.~R.} 2012, 
\mnras, 426, 2751 

\end{thebibliography}
\end{document}